\documentclass[twocolumn,iop,numberedappendix,appendixfloats]{oja}

\usepackage[utf8]{inputenc}
\usepackage[british]{babel}

\usepackage{graphicx}	
\usepackage{amsmath}	
\usepackage{amssymb}	
\usepackage{subcaption}  
\usepackage{caption}
\usepackage{booktabs}

\usepackage{xcolor}
\usepackage{soul}
\definecolor{maroon}{cmyk}{0, 0.87, 0.68, 0.32}
\definecolor{halfgray}{gray}{0.55}
\definecolor{ipython_frame}{RGB}{207, 207, 207}
\definecolor{ipython_bg}{RGB}{247, 247, 247}
\definecolor{ipython_red}{RGB}{186, 33, 33}
\definecolor{ipython_green}{RGB}{0, 128, 0}
\definecolor{ipython_cyan}{RGB}{64, 128, 128}
\definecolor{ipython_purple}{RGB}{170, 34, 255}
\usepackage{float}
\usepackage{fontawesome}
\usepackage[colorlinks,allcolors=blue]{hyperref}
\usepackage{url}

\usepackage{array,multirow,graphicx}

\usepackage{graphicx}	
\usepackage{amsmath}	
\usepackage{booktabs}   

\newcommand{\lcdm}{$\Lambda$CDM}
\newcommand{\wcdm}{$w$CDM}
\newcommand{\wowacdm}{$w_0 w_a$CDM}
\newcommand{\prob}{\ensuremath{{p}}}
\newcommand{\data}{\ensuremath{\boldsymbol{{d}}}}

\newcommand{\software}[1]{\footnote{\protect\url{#1}}}

\newcommand{\lik}{{\mathcal{L}}}

\newcommand{\diff}{{\rm{d}}}
\newcommand{\vect}[1]
  {\mbox{\boldmath ${#1}$}}

\newcommand{\parsone}
  {\vect{\theta}}
\newcommand{\parstwo}
  {\vect{\eta}}
\newcommand{\bayesfactor} 
  {\frac{z_1}{z_2}}

\usepackage{listings}

\lstdefinelanguage{iPython}{
    morekeywords={access,and,break,class,continue,def,del,elif,else,except,exec,finally,for,from,global,if,import,in,is,lambda,not,or,pass,print,raise,return,try,while},%
    morekeywords=[2]{abs,all,any,basestring,bin,bool,bytearray,callable,chr,classmethod,cmp,compile,complex,delattr,dict,dir,divmod,enumerate,eval,execfile,file,filter,float,format,frozenset,getattr,globals,hasattr,hash,help,hex,id,input,int,isinstance,issubclass,iter,len,list,locals,long,map,max,memoryview,min,next,object,oct,open,ord,pow,property,range,raw_input,reduce,reload,repr,reversed,round,set,setattr,slice,sorted,staticmethod,str,sum,super,tuple,type,unichr,unicode,vars,xrange,zip,apply,buffer,coerce,intern},%
    sensitive=true,%
    morecomment=[l]\#,%
    morestring=[b]',%
    morestring=[b]",%
    morestring=[s]{'''}{'''},
    morestring=[s]{"""}{"""},
    morestring=[s]{r'}{'},
    morestring=[s]{r"}{"},%
    morestring=[s]{r'''}{'''},%
    morestring=[s]{r"""}{"""},%
    morestring=[s]{u'}{'},
    morestring=[s]{u"}{"},%
    morestring=[s]{u'''}{'''},%
    morestring=[s]{u"""}{"""},%
    literate=
    {á}{{\'a}}1 {é}{{\'e}}1 {í}{{\'i}}1 {ó}{{\'o}}1 {ú}{{\'u}}1
    {Á}{{\'A}}1 {É}{{\'E}}1 {Í}{{\'I}}1 {Ó}{{\'O}}1 {Ú}{{\'U}}1
    {à}{{\`a}}1 {è}{{\`e}}1 {ì}{{\`i}}1 {ò}{{\`o}}1 {ù}{{\`u}}1
    {À}{{\`A}}1 {È}{{\'E}}1 {Ì}{{\`I}}1 {Ò}{{\`O}}1 {Ù}{{\`U}}1
    {ä}{{\"a}}1 {ë}{{\"e}}1 {ï}{{\"i}}1 {ö}{{\"o}}1 {ü}{{\"u}}1
    {Ä}{{\"A}}1 {Ë}{{\"E}}1 {Ï}{{\"I}}1 {Ö}{{\"O}}1 {Ü}{{\"U}}1
    {â}{{\^a}}1 {ê}{{\^e}}1 {î}{{\^i}}1 {ô}{{\^o}}1 {û}{{\^u}}1
    {Â}{{\^A}}1 {Ê}{{\^E}}1 {Î}{{\^I}}1 {Ô}{{\^O}}1 {Û}{{\^U}}1
    {œ}{{\oe}}1 {Œ}{{\OE}}1 {æ}{{\ae}}1 {Æ}{{\AE}}1 {ß}{{\ss}}1
    {ç}{{\c c}}1 {Ç}{{\c C}}1 {ø}{{\o}}1 {å}{{\r a}}1 {Å}{{\r A}}1
    {€}{{\EUR}}1 {£}{{\pounds}}1
    {^}{{{\color{ipython_purple}\^{}}}}1
    {=}{{{\color{ipython_purple}=}}}1
    {+}{{{\color{ipython_purple}+}}}1
    {-}{{{\color{ipython_purple}-}}}1
    {*}{{{\color{ipython_purple}$^\ast$}}}1
    {/}{{{\color{ipython_purple}/}}}1
    {+=}{{{+=}}}1
    {-=}{{{-=}}}1
    {*=}{{{$^\ast$=}}}1
    {/=}{{{/=}}}1,
    literate=
    *{-}{{{\color{ipython_purple}-}}}1
     {?}{{{\color{ipython_purple}?}}}1,
    identifierstyle=\color{black}\ttfamily,
    commentstyle=\color{ipython_cyan}\ttfamily,
    stringstyle=\color{ipython_red}\ttfamily,
    keepspaces=true,
    showspaces=false,
    showstringspaces=false,
    rulecolor=\color{ipython_frame},
    frameround={t}{t}{t}{t},
    numbers=none,
    numberstyle=\tiny\color{halfgray},
    backgroundcolor=\color{ipython_bg},
    basicstyle=\ttfamily\footnotesize,
    columns=fullflexible,
    keywordstyle=\color{ipython_green}\ttfamily,
}

\begin{document}

\journalinfo{The Open Journal of Astrophysics}
\submitted{submitted XXX; accepted YYY}

\title[SDDR with normalizing flows]{Savage-Dickey density ratio estimation with normalizing flows\\ for Bayesian model comparison}\vspace{-10ex}
\shorttitle{SDDR with normalizing flows}
\shortauthors{Lin et al.}

\author{
	Kiyam Lin,$^{1\star}$\thanks{E-mail: kiyam.lin@ucl.ac.uk}
	Alicja Polanska,$^{1}$
	Davide Piras,$^{2,3}$
	Alessio Spurio Mancini,$^{4}$
	and Jason D. McEwen$^{1}$}

\affiliation{$^1$ Mullard Space Science Laboratory, University College London, Holmbury St. Mary, Dorking, Surrey, RH5 6NT, UK}
\affiliation{$^2$ Centre Universitaire d’Informatique, Université de Genève, 7 route de Drize, 1227 Genève, Switzerland}
\affiliation{$^3$ Département de Physique Théorique, Université de Genève, 24 quai Ernest Ansermet, 1211 Genève 4, Switzerland}
\affiliation{$^4$ Department of Physics, Royal Holloway, University of London, Egham Hill, Egham, UK}

\thanks{$^\star$ E-mail: \href{mailto:kiyam.lin@ucl.ac.uk}{kiyam.lin@ucl.ac.uk}}

\date{\today}

\begin{abstract}
	A core motivation of science is to evaluate which scientific model best explains observed data.
	Bayesian model comparison provides a principled statistical approach to comparing scientific models and has found widespread application within cosmology and astrophysics.
	Calculating the Bayesian evidence is computationally challenging, especially as we continue to explore increasingly more complex models.
	The Savage-Dickey density ratio (SDDR) provides a method to calculate the Bayes factor (evidence ratio) between two nested models using only posterior samples from the super model.
	The SDDR requires the calculation of a normalised marginal distribution over the extra parameters of the super model, which has typically been performed using classical density estimators, such as histograms.
	Classical density estimators, however, can struggle to scale to high-dimensional settings.
	We introduce a neural SDDR approach using normalizing flows that can scale to settings where the super model contains a large number of extra parameters.
	We demonstrate the effectiveness of this neural SDDR methodology applied to both toy and realistic cosmological examples.
	For a field-level inference setting, we show that Bayes factors computed for a Bayesian hierarchical model (BHM) and simulation-based inference (SBI) approach are consistent, providing further validation that SBI extracts as much cosmological information from the field as the BHM approach.
	The SDDR estimator with normalizing flows is implemented in the open-source \texttt{harmonic} Python package. \href{https://github.com/astro-informatics/harmonic/}{\faGithub}
\end{abstract}

\maketitle




\section{Introduction}
A core aspect of science is to determine which scientific model of reality is the most accurate.
Bayesian model comparison provides a principled statistical approach to comparing scientific models that has found widespread application, particularly in astrophysics and cosmology \citep{Trotta07}.
With advances in computational power, it is possible to analyse increasingly complex cosmological and astrophysical models.
To describe these more complex and intricate models, additional parameters are often required, including those that are physically interesting as well as those that might be considered nuisance parameters yet must be modelled \citep[e.g.][]{mvwk2024kids}.

These more complex models that often exist in higher-dimensional parameter spaces, however, can be costly to analyse, with some models being infeasible to explore due to insurmountable computational costs. We can make use of modern machine learning techniques and their underlying technologies to massively accelerate the computational speed of performing Bayesian inference. This can be achieved by combining (i) \textit{emulation of physical models by machine learning techniques}; (ii) \textit{differentiable and probabilistic programming}; (iii) \textit{scalable Markov chain Monte Carlo sampling} (MCMC) that makes use of gradient information, and (iv) \textit{decoupled and scalable Bayesian model selection}, as advocated in \citet{piras2024future}. Gradient-based Markov chain Monte Carlo (MCMC) sampling algorithms that can leverage gradients computed efficiently by automatic differentiation make sampling in higher-dimensional parameter spaces tractable, e.g.\ the No-U-Turn Sampler (NUTS; \citealt{Hoffman14}), a highly efficient and adaptive variant of Hamiltonian Monte Carlo (HMC; \citealt{Duane87, Neal96}).

Bayesian model selection requires the computation of the Bayesian evidence, which is typically performed by nested sampling (\citealt{Skilling06}; see \citealt{ashton22nested, buchner21nested} for reviews), where sampling must be performed in a prescribed nested manner for which a variety of algorithms have been developed \citep{, Feroz08, Feroz09, higson2019dynamic, Feroz19, Handley15, Handley15b, speagle2020dynesty, Buchner21, cai2022proximal, mcewen2023proximal, lange2023nautilus}. Nested sampling is thus not applicable for arbitrary MCMC sampling algorithms. To compute the Bayesian evidence for arbitrary sampling algorithms, such as NUTS, we require methods to compute the Bayesian evidence that are decoupled from the sampling algorithm while being scalable. Methods that require only posterior samples to compute the evidence have become increasingly popular \citep{Dickey71, diciccio1997computing, Trotta07, heavens2017marginal, Jia19, mcewen2021machine, Srinivasan24, Rinaldi24}. Amongst these methods, the learned harmonic mean estimator \citep{mcewen2021machine, Polanska23, polanska2024learned}, available in the \texttt{harmonic} software package, has been shown to provide robust and scalable evidence calculations and has been growing in popularity within astrophysics and cosmology \citep{spurio2023bayesian, piras2024future, polanska2024flowmc, mancini2024field, stiskalek2024symmetry, du2025desi, carrion2025shear, stiskalek2025velocity}.

Even with such acceleration, the process of performing Bayesian model comparison can still be highly computationally costly as it requires acquiring posterior samples along with their respective likelihood values for every model being evaluated. However, if the models being compared have a relationship where one model is a \textit{nested} model of the larger \textit{super} model, then we can make use of the Savage-Dickey density ratio (SDDR; \citealt{Dickey71, verdinelli1995computing, o1999kendall}) to perform Bayesian model comparison without needing to explicitly acquire samples from the nested model. The term `nested' here refers to a relationship between two models where a super model reduces to a smaller, nested model when some of the super model's parameters are fixed at particular values, as is often the case in cosmological settings. We refer to the parameters common to both the super and nested model as \textit{common parameters} and the additional parameters of the super model as \textit{extra parameters}. This approximately halves the computational cost of performing Bayesian model comparison as we only need to perform sampling once for the super model. The SDDR has already been applied to cosmology, for instance by \citet{Trotta07, marin2010resolving, verde2013lack, leistedt2014no, salvatelli2014indications, di2017can, di2020planck,kreisch2020neutrino}.

Besides requiring nested models, the SDDR also requires there to be sufficient posterior support with high enough sample density around the parameter values where the super model reduces to the nested model to be able to evaluate the marginal posterior accurately. If this requirement is not met, calculating the evidence directly with the learned harmonic mean provides a good alternative.

Traditionally the SDDR has been calculated through the use of classical density estimators, such as histograms or kernel density estimation. For situations where there are only a small number of extra parameters, classical density estimators provide good estimates of the evidence ratio. However, this approach can fail even in settings with just a few extra parameters as, e.g., it becomes computationally difficult to normalise the histogram of the probability distribution of the extra parameters by numerical integration due to the curse of dimensionality. Normalizing flows \citep{papamakarios2021normalizing}, however, are normalized by construction and thus are a natural neural alternative to classical density estimators when considering higher-dimensional problems. Thus, in this work we make use of normalizing flows to estimate the normalized probability distribution of the extra parameters. As such, we are able to perform SDDR calculations with a large number of extra parameters.

The remainder of this article is organised as follows. In Section \ref{sec:sddr} we provide a brief overview of SDDR and our methodology with normalizing flows. In Section \ref{sec:applications} we demonstrate and validate the use of our SDDR methodology in both toy and realistic cosmological examples. We conclude in Section \ref{sec:conclusions}.

\section{Savage-Dickey density ratio with normalizing flows} \label{sec:sddr}

In this section we provide a brief overview of the SDDR and its connection to Bayesian model comparison. We introduce our SDDR methodology with normalizing flows, discuss how uncertatinties can be estimated and present an open-source software implementation.

\subsection{Bayesian model comparison}

The Bayesian evidence is a measure of the probability of obtaining the observed data given the model under consideration, defined as

\begin{equation} \label{eq:evidence}
	z_i = p(\data | M_i) = \int{p(\data | \theta, M_i) p(\theta | M_i) \diff \theta},
\end{equation}

\noindent where $z_i$ denotes the Bayesian evidence of model $M_i$ for data, $\data$. The evidence ratio, or Bayes factor, between any two models, $M_1$ and $M_2$, can be used to assess which model is favored by the data, expressed as

\begin{equation} \label{eq:evidence_ratio}
	\bayesfactor = \frac{\prob(\data | M_1)}{\prob(\data | M_2)}.
\end{equation}

\noindent In the case where we have equal prior probabilities for each model, as is often the case in practice, the Bayes factor reduces to the posterior model odds following Bayes theorem:

\begin{equation} \label{eq:posterior_odds}
	\bayesfactor = \frac{p(M_1 | \data) p(M_2)}{p(M_2 | \data) p(M_1)} = \frac{\prob(M_1 | \data)}{\prob(M_2 | \data)}.
\end{equation}

\noindent For models with equal prior model probabilities, the Bayes factor can thus be used to inform us of which model is more probable given the observed data.

\subsection{Savage-Dickey density ratio}

The SDDR \citep{Dickey71, verdinelli1995computing, o1999kendall} is a method of calculating the Bayes factor, i.e. ratio of evidences between two models, when one model is a nested model of the other. It is useful when one does not have direct access to the nested model or it is computational infeasible to explore all possible nested models. The SDDR gives the Bayes factor in terms of the ratio of the normalized marginal posterior to the prior density of the super model in the extra parameter(s). This ratio is evaluated at the point where the super model reduces to the nested model. Model $M_1$ can be considered a nested model of a larger super model $M_2$ if:

\begin{enumerate}
	\item the parameters of $M_2$ can be expressed as a set of parameters $\{\parsone, \parstwo\}$, where $\parsone$ are the common parameters of $M_1$ and $M_2$, whilst $\parstwo$ are the extra parameters required to extend $M_1$ to $M_2$;
	\item for some particular value of $\parstwo=\parstwo_1$, the likelihood of the super model reduces to the likelihood of the nested model;
	\item the prior distributions of \parsone\ are identical for $M_1$ and $M_2$;
	\item the prior distributions of \parstwo\ and \parsone\ are separable, i.e. $\prob(\parsone, \parstwo) = \prob(\parsone) \prob(\parstwo)$.
\end{enumerate}

The Bayes factor given data and two models $M_1$ and $M_2$ is given by
\begin{eqnarray}
	\bayesfactor & = & \frac{\prob(\data | M_1)}{\prob(\data | M_2)} \nonumber \\
	& = & \frac{
		\int \prob(\data | \parsone^{\prime}, M_1) \,
		\prob(\parsone^{\prime} | M_1) \,
		\diff \parsone^{\prime}
	}
	{
		\int \prob(\data | \parsone^{\prime\prime}, \parstwo^{\prime\prime}, M_2) \,
		\prob(\parsone^{\prime\prime}, \parstwo^{\prime\prime} | M_2) \,
		\diff \parsone^{\prime\prime} \,
		\diff \parstwo^{\prime\prime}
	}. \label{equation:bf_sddr}
\end{eqnarray}

As our models are nested, we have $\prob(\data | \parsone, M_1) = \prob(\data | \parsone, \parstwo_1, M_2)$. It is then straightforward to show \citep{Dickey71, verdinelli1995computing, o1999kendall, Trotta07, verde2013lack} that the Bayes factor reduces to

\begin{equation}
	\bayesfactor =\frac{
		\int \prob(\parsone^\prime, \parstwo_1 | \data, M_2) \,
		\diff \parsone^\prime
	}
	{ \prob( \parstwo_1 | M_2) }.
\end{equation}

We can see that the integration of the posterior distribution depicted in the numerator is simply marginalization over the common parameters \parsone, yielding the marginalized posterior in $\parstwo_1$. Hence, the Bayes factor between two nested models is given by

\begin{equation}
	\label{equation:sddr}
	\bayesfactor
	= \frac{\prob(\parstwo_1 | \data, M_2)}{\prob(\parstwo_1 | M_2)}.
\end{equation}

The numerator in Eq. \eqref{equation:sddr} is the marginalised posterior of the extra parameter(s) for the super model, whilst the denominator is the prior of the extra parameter(s) for the super model. A more detailed derivation of the SDDR can be found in Appendix \ref{sec:appendix}.

It follows from Eq. \eqref{equation:sddr} that to compute the SDDR one only requires the posterior samples from the super model and the prior distribution of the extra parameter(s). This is a significant advantage as it means that we only need to perform sampling once for the super model. We can then use those samples to calculate the Bayes factor with relation to all possible nested models.

As evaluation of the prior is typically trivial, the challenge lies in the evaluation of the marginal posterior at $\parstwo_1$, as it must be normalized. This requires numerical integration in potentially high-dimensions if there are many extra parameters. In this work, we consider two approaches.

In the first approach we make use of classical density estimators by way of a normalized histogram (as considered previously, e.g. \citealt{verde2013lack}). We also introduce a second approach where we make use of a neural density estimator by way of normalizing flows to learn the marginal posterior distribution. Importantly, it is only with normalizing flows that we are able to calculate the SDDR in high-dimensional marginal posterior spaces.

Care however needs to be taken when using the SDDR as it is only valid when the posterior has sufficient sample density around $\parstwo_1$. If this requirement is not met, calculating the evidence directly with the learned harmonic mean is preferred.

\subsection{SDDR with classical density estimators}

To compute the SDDR with classical density estimators we consider a simple normalized histogram. Following normalization, we evaluate the marginal posterior probability at $\parstwo_1$. Beyond even a few dimensions normalizing a histogram becomes computationally challenging. This is due to the need to numerically integrate over the high-dimensional parameter space alongside the curse of dimensionality. Consequently, this method does not scale well to high-dimensional problems.

\subsection{SDDR with normalizing flows}

Normalizing flows have the advantage of being normalized by construction, and thus are a natural neural density estimator analogue to classical density estimators that scale to higher-dimensional settings. A normalizing flow, in essence, learns the transformation from a known tractable base distribution such as a Gaussian to the target distribution. For a comprehensive review of normalizing flows we refer the reader to \citet{papamakarios2021normalizing}.

More specifically, a vector $\vect{\theta}$ from an unknown distribution $p(\vect{\theta})$ can be expressed through a transformation $B$ of a latent vector $\vect{u}$ sampled from the base distribution $q(\vect{u})$:
\begin{equation}
	\vect{\theta} = B(\vect{u}), \quad \vect{u}	 \sim q(\vect{u}). \label{eq:transformation}
\end{equation}
If $B$ is invertible and both $B$ and its inverse $B^{-1}$ are differentiable, we can calculate the density of the distribution of $\vect{\theta}$ through the change of variables formula by
\begin{equation}
	p(\vect{\theta}) = q(\vect{u}) | \det J_B(\vect{u}) |^{-1}, \label{eq:change_of_variables}
\end{equation}
where $J_B(\vect{u})$ is the Jacobian corresponding to $B$. As this transformation is composable, we can apply $B$ multiple times to obtain approximations of more complex target distributions. A normalizing flow thus consists of a series of such transformations.

In this work we make use of rational quadratic spline flows \citep{durkan2019neural}. Rational quadratic spline flows split the input domain of $\vect{u}$ into multiple segments and apply monotonic rational quadratic polynomial transformations to each segment. The polynomial segments are restricted to match at bin boundaries such that the transformed distribution is continuous and differentiable.

The rational quadratic splines consist of learned parameters that are optimised during training as well as alternating affine transformation layers to create the normalizing flow. Network training is performed by minimizing the Kullback-Leibler divergence between samples of the marginal posterior and the target distribution parameterised by the normalizing flow. We train the flow to learn the marginal posterior distribution and evaluate the flow at $\parstwo_1$.

\subsection{Uncertainty estimation}

To obtain error estimates for both the classical and neural SDDR methodologies we perform bootstrapping. We randomly subsample the marginalised posterior samples, with replacement, to create several sets of bootstrapped samples. We then make use of each bootstrapped sample set to calculate an SDDR estimate either with an independently trained flow or a normalized histogram. We take the mean of these estimates as our SDDR estimate and the standard deviation across estimates as the error estimate.

\subsection{Code implementation}

The implementation of the SDDR using normalizing flows can be found within the \texttt{harmonic}\software{https://github.com/astro-informatics/harmonic} software package.
We have chosen to implement our methods as part of \texttt{harmonic} to provide a single code for Bayesian model comparison that is versatile and can handle all situations. The implementation is also consistent with the interface of \texttt{harmonic}. \texttt{harmonic} is a professional code that is well-tested and documented and can be installed from PyPi.
Given posterior samples, computation of the Bayes factor using our implementation of the SDDR using normalizing flows requires very little computational resources to train and run, typically running within a few minutes on a single CPU. In the snippet in listing \ref{lst:sddr_example} we demonstrate the use of SDDR within harmonic.

\begin{lstlisting}[language=iPython, mathescape=true, caption={Example code snippet of how to use the SDDR functionality within \texttt{harmonic} to calculate the Bayes factor given $\parstwo_1$ and its corresponding log prior value as well as the samples of the extra parameters.}, captionpos=b, label=lst:sddr_example] 
import harmonic as hm
import numpy as np

# Define the number of marginal dimensions
nparams = 4
# Load the marginalised posterior samples
marginalised_samples = np.load('samples.npy') 
# Define the pre-calculated log prior value
log_prior_$\eta$_1 = -2
# The point of nesting
$\eta$_1 = np.array([-2.0, 0.0, -0.5, 0.5])
# Create the normalizing flow model
model = hm.model.RQSplineModel(nparams, temperature=1.0)
# Create the sddr object
sddr = hm.sddr.sddr(model, marginalised_samples)
# Calculate the Bayes factor
log_bf, std = sddr.log_bayes_factor(log_prior_$\eta$_1, $\eta$_1)
\end{lstlisting}

\section{Applications} \label{sec:applications}

We calculate the SDDR with both classical density estimators (histogram) and neural density estimators (normalizing flows) for a variety of toy and realistic cosmological examples. Furthermore, we validate the Bayes factors computed against alternative approaches such as nested sampling and the learned harmonic mean (which require sampling both the super and nested models), finding excellent agreement. We demonstrate the effectiveness of neural SDDR both in low-dimensional marginal posterior spaces where we have access to the classical density estimator method, but also in higher-dimensional extra parameter spaces where the classical density estimator method breaks down.

\subsection{Gaussian toy examples} \label{sec:gaussian_toy}

We construct two toy Gaussian likelihood examples with either 1 or 4 extra parameters in the super model compared to the nested model. In all scenarios we calculate the log Bayes factor (evidence ratio) by the SDDR, making use of both the classical histogram method and the normalizing flow method. We validate our results with a combination of both the learned harmonic mean estimator using \texttt{harmonic} with samples obtained with a MCMC sampler via \texttt{emcee}\software{https://github.com/dfm/emcee} \citep{foreman2013emcee}, and nested sampling with the use of \texttt{nautilus}\software{https://github.com/johannesulf/nautilus} \citep{lange2023nautilus} to have two independent methods of calculating the Bayesian evidence.

We consider the data model $\data \sim \mathcal{N}(\vect{\mu}(\vect{\theta}), \Sigma)$, where the mean $\vect{\mu}$ is dependent on parameters $\vect{\theta}$ but the covariance matrix $\Sigma$ is not.
The corresponding Gaussian likelihood is given by
\begin{equation}
	\lik( \data \mid \vect{\theta}) = \frac{\exp\bigl( -\frac{1}{2} (\data - \vect{\mu}(\vect{\theta}))^\text{T} \Sigma^{-1} (\data - \vect{\mu}(\vect{\theta}) \bigr)}{(2 \pi)^{n/2}|\Sigma|^{1/2}}, \label{eq:gaussian_likelihood}
\end{equation}
where $n$ is the dimension of the data vector.

For the first Gaussian example we consider a data dimension of $n=3$ and a super model with parameter dimension $d=4$ where
\begin{align}
	\vect{\mu}(\vect{\theta}) & = \begin{bmatrix}
		                              \theta_1 \\
		                              \theta_2 \\
		                              e^{0.5 \times \theta_3} + \theta_4
	                              \end{bmatrix}.
\end{align}
In this first Gaussian model considered, the super model (with 4 parameters) reduces to the nested model with 3 parameters when $\theta_4 = -2.0$. We initialise a random covariance, $\Sigma$, by first creating a diagonal covariance with elements randomly drawn from a 1D Gaussian with mean of $1.0$ and standard deviation of $0.1$ ensuring all elements are positive. Off diagonal elements in the upper triangle that are only one index away from the diagonal with matrix index $i$ are then set with $\Sigma_{i, i+1} = -1 ^{i} \times 0.5 \times \sqrt{\Sigma_{i,i} \times \Sigma_{i+1,i+1}}$. The upper triangle is then symmetrised with the lower triangle making the covariance matrix symmetrical. A mock data vector is generated by evaluating the forward model at $\vect{\theta} = [0.0, -0.5, 0.5, -2.0]$, making the nested model the ground truth.  For inference we consider Gaussian priors on all parameters with mean of 0.0 and diagonal covariance of 2.0.

We find that with this simple example both the histogram and normalizing flow methods provide accurate estimates of the Bayes factor that are in close agreement with those computed by nested sampling and the learned harmonic mean, as shown in Table \ref{tab:1d_gaussian}. In this case, there is a weak preference for the nested model which is indeed the model used to generate the ground truth.

\begin{table}
	\centering
	\caption{Gaussian toy model with 1 extra parameter.}
	\begin{tabular} {l c}
		\toprule
		Method                & Log Bayes factor  \\
		\midrule
		Nested sampling       & $0.905 \pm 0.013$ \\
		Learned harmonic mean & $0.877 \pm 0.002$ \\
		SDDR (classical)      & $0.893 \pm 0.021$ \\
		SDDR (flows)          & $0.890 \pm 0.056$ \\
		\bottomrule
	\end{tabular}
	\label{tab:1d_gaussian}
\end{table}

For the second toy Gaussian example, we consider the same setup as before but we have a super model that has 8 parameters whilst the nested model only has 4 parameters. The likelihood is again of the form shown in Eq.\ \ref{eq:gaussian_likelihood}, this time with $n=4$, and the super model is defined with
\begin{align}
	\vect{\mu}(\vect{\theta}) & = \begin{bmatrix}
		                              \theta_1 + \operatorname{arcsinh}(\theta_5)  \\
		                              \operatorname{arctan}(\theta_2) + \theta_6^2 \\
		                              e^{0.5 \times \theta_2} + \theta_7           \\
		                              e^{0.5 \times \theta_4} + \operatorname{arctan}(\theta_8)
	                              \end{bmatrix},
\end{align}
\noindent which reduces to the nested model by setting $\theta_5 = -2.0$, $\theta_6 = 0.2$, $\theta_7 = -0.2$ and $\theta_8 = 1.5$. A mock data vector is generated by evaluating the forward model at $\vect{\theta} = [0.0, -0.5, 0.5, 1.0, -2, 0.2, -0.2, 1.5]$, making the nested model the ground truth once again.  For inference we again consider Gaussian priors on all parameters with mean of 0.0 and diagonal covariance of 2.0.

\begin{figure*}
	\centering
	\includegraphics[width=\textwidth]{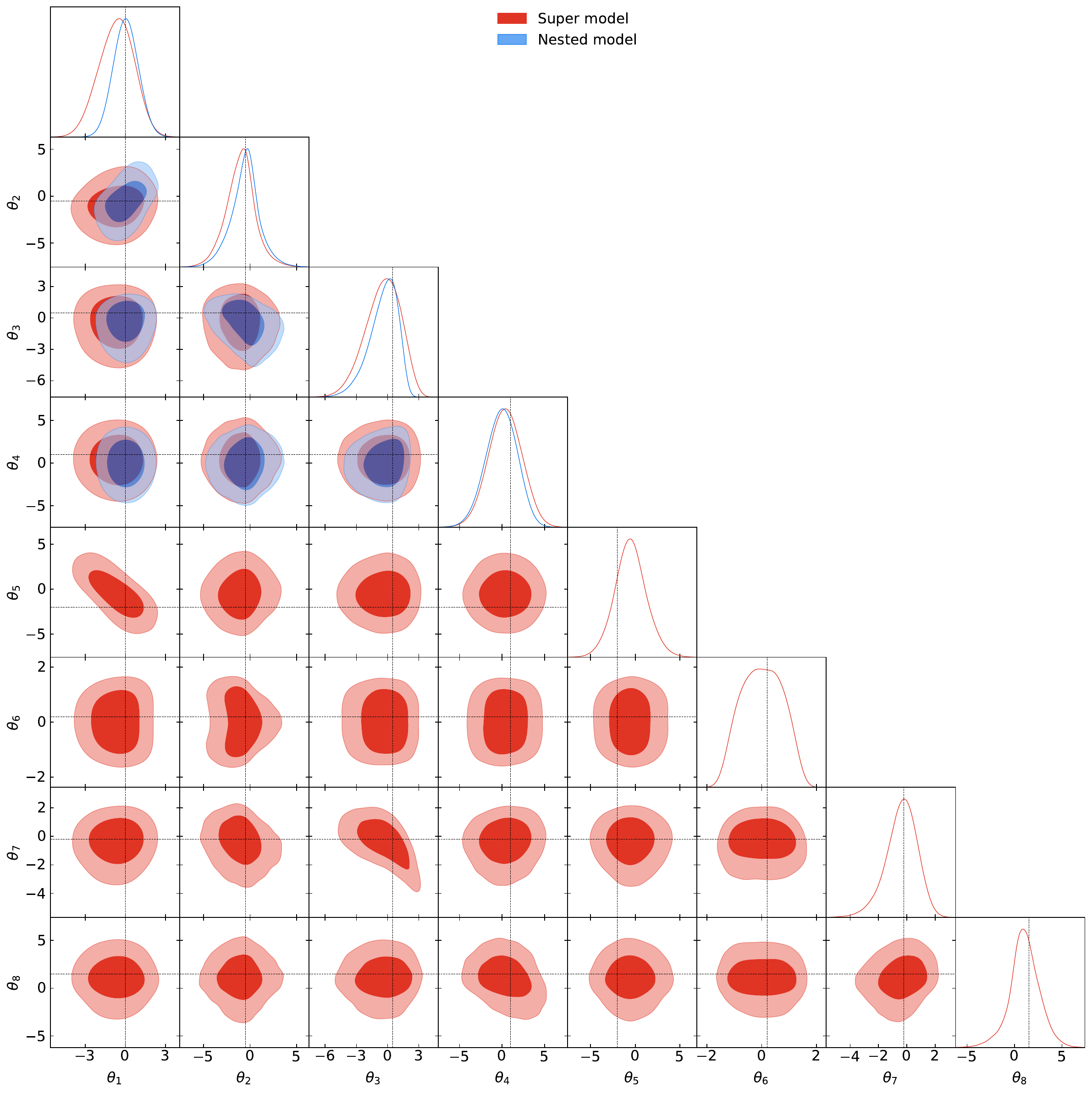}
	\caption{Posterior contours of the toy Gaussian model with 4 extra parameters including both the super model (red) and the nested model (blue). The dashed lines depict the ground truth.}
	\label{fig:4d_gaussian}
\end{figure*}

Already in this setting we found it computationally challenging to normalize the histogram accurately. As a result we are unable to easily compute the SDDR with the histogram methodology by direct numerical integration. Instead, we rely on the normalizing flow methodology to calculate SDDR and find that it is in good agreement with the Bayes factor calculated from both nested sampling and the learned harmonic mean as shown in Table \ref{tab:4d_gaussian}. Fig. \ref{fig:4d_gaussian} depicts the posterior contours with truth markers for both the nested and super model and the Bayes factor indicates that there is a definitive preference for the nested model, which is the model used to generate the ground truth.

\begin{table}
	\centering
	\caption{Gaussian toy model with 4 extra parameters. Note that there is no entry for the histogram method as already in this simple example it was computationally challenging to normalize the marginal posterior histogram and thus calculate the SDDR.}
	\begin{tabular} {l c}
		\toprule
		Method                & Log Bayes factor  \\
		\midrule
		Nested sampling       & $2.227 \pm 0.013$ \\
		Learned harmonic mean & $2.248 \pm 0.007$ \\
		SDDR (classical)      & ---               \\
		SDDR (flows)          & $2.374 \pm 0.088$ \\
		\bottomrule
	\end{tabular}
	\label{tab:4d_gaussian}
\end{table}

Given that the toy Gaussian examples show good agreement between the Bayes factor calculated with the classical and flow SDDR methodologies, nested sampling and the learned harmonic mean, we will now turn our attention to more realistic cosmological examples.

\subsection{Dark Energy Survey: \lcdm\ vs. \wcdm}

The first more realistic cosmological scenario explored is with the Dark Energy Survey (DES; \citealt{dark2005dark}). We consider simulated DES Year 1 weak lensing and clustering data following the approach described in \citet{campagne2023jax}.

For this example two models are considered. The super model is the \wcdm\ cosmological model where dark energy is treated as a fluid with equation of state characterised by the $w$ parameter. The corresponding nested model is \lcdm, which corresponds to setting $w=-1$ in the \wcdm\ model. In this work we directly make use of the posterior \wcdm\ samples obtained already by \citet{polanska2024learned} by MCMC sampling.  We also compare to the Bayes factors calculated by \citet{polanska2024learned} using the learned harmonic mean and nested sampling.

Table \ref{tab:DES} reports the results obtained by our SDDR methodology for the DES cosmology analysis. There is a definitive preference for \lcdm\ matching the truth. We find that our methodology is in good agreement with the results obtained in the original analysis. A clear advantage here is that we are able to obtain these evidence ratios without needing to perform additional MCMC sampling of the nested \lcdm\ model, approximately halving the computational cost as the SDDR method has negligible computational cost compared to the MCMC sampling of the models.

\begin{table}
	\centering
	\caption{DES cosmology \lcdm\ vs. \wcdm\ showing correct preference for \lcdm\ given mock ground truth.}
	\begin{tabular} {l c}
		\toprule
		Method                & Log Bayes factor \\
		\midrule
		Nested sampling       & $2.23 \pm 0.45$  \\
		Learned harmonic mean & $2.15 \pm 0.01$  \\
		SDDR (classical)      & $2.14 \pm 0.02$  \\
		SDDR (flows)          & $2.13 \pm 0.02$  \\
		\bottomrule
	\end{tabular}
	\label{tab:DES}
\end{table}

\subsection{Stage-IV weak lensing: \lcdm\ vs. \wowacdm}

We explore another more realistic cosmological example that was also considered in \citet{piras2024future}. The models being compared are those of \lcdm\ and \wowacdm. The super model of \wowacdm\ is a phenomenological model where dark energy is not only treated as a fluid, but is also allowed to evolve with time, characterised by the two parameters of $w_0$ and $w_a$. It reduces to \lcdm\ when we set $w_0 = -1$ and $w_a = 0$. This analysis followed a next-generation simulated weak lensing setup and featured a super model that had 39 parameters and a corresponding nested model with 37 parameters.

Such higher-dimensional spaces are typically more challenging to explore. If one is only interested in the Bayes factor, however, then the SDDR can alleviate challenges presented from direct evidence calculation as we only need concern ourselves with the extra parameters of the super model, \parstwo, and the evaluation of that marginal posterior volume at $\parstwo_1$.

Table \ref{tab:future} shows the results obtained through the classical density (histogram) and neural (normalizing flow) SDDR methodologies. The results we obtain with the SDDR are broadly consistent with the results obtained in \citet{piras2024future}, showing correct preference for the \lcdm\ model used to generate the ground truth.

\begin{table}
	\centering
	\caption{Stage-IV weak lensing \lcdm\ vs. \wowacdm\ showing correct preference for \lcdm\ given mock ground truth.}
	\begin{tabular} {l c}
		\toprule
		Method                & Log Bayes factor \\
		\midrule
		Nested sampling       & $0.78 \pm 0.71$  \\
		Learned harmonic mean & $1.53 \pm 0.07$  \\
		SDDR (classical)      & $1.75 \pm 0.06$  \\
		SDDR (flows)          & $1.75 \pm 0.06$  \\
		\bottomrule
	\end{tabular}
	\label{tab:future}
\end{table}

\subsection{Field-level weak lensing inference: \lcdm\ vs. \wcdm}

In the field-level setting, where the evidence is typically challenging to compute due to the high-dimensionality of parameters, we can still make use of the SDDR to perform model comparison. The SDDR methodology is only dependent on the marginal posterior samples of parameter $\parstwo$\ and the ability to evaluate the posterior probability at the point of nesting, i.e. at $\parstwo = \parstwo_1$, and so even in high-dimensional inference settings can provide a robust and precise estimate of the Bayes factor.

For the field-level setting, we consider the \wcdm\ analysis presented in \citet{lanzieri2024optimal, zeghal2024simulation}. This scenario considers \wcdm\ with a field-level cosmological analysis constructed using a Bayesian hierarchical model (BHM) with a mock stage-IV weak lensing survey setup. Like the DES Y1 example presented previously, the \wcdm\ model reduces to the \lcdm\ model when setting $w=-1$. This is in general a costly setup to run even when harnessing modern hardware and gradient-accelerated methodologies, with \citet{zeghal2024simulation} writing that the explicit full field inference required $\mathcal{O}(10^5)$\ to $\mathcal{O}(10^6)$ forward model evaluations to converge.

Previous works \citep{lanzieri2024optimal, zeghal2024simulation} did not explicitly calculate the Bayesian evidence. However, using the SDDR approach we are able to calculate the Bayes factor for the BHM setting between the \wcdm\ and \lcdm\ models for the first time. As we do not have access to the field-level evidence either calculated with the learned harmonic mean or with nested sampling directly for the BHM setting, we validate our results by comparison with a simulation-based inference (SBI) analysis that compressed the field with a convolutional neural network. We leverage recent work presented in \citet{mancini2024field} which made use of the same simulator setup as in \citet{lanzieri2024optimal} but performed an SBI analysis based on neural likelihood estimation (in contrast to the neural posterior estimation performed by \citealt{lanzieri2024optimal}). We make use of the SBI chains computed by \citet{mancini2024field} to calculate the log Bayes factor with the SDDR.

Table \ref{tab:sbi_lens} shows the results obtained from the SDDR methodology for the field-level BHM and SBI \wcdm\ analysis with both classical (histogram) and neural (normalizing flows) methods, showing consistent results and correct preference for the \lcdm\ model used to generate the ground truth.
We compare the Bayes factors computed by the SDDR for the SBI scenario to that computed by the learned harmonic mean estimator, finding close agreement with respect to the interpretation of the Bayes factor on the Jeffreys scale \citep{jeffreys1998theory, nesseris2013jeffrey} leading to the same conclusions.  Moreover, the Bayes factors computed by the BHM and SBI approaches are also in close agreement, further validating the accuracy of not only the SDDR approach but also the SBI approximate inference framework.

\begin{table}
	\centering
	\begin{tabular} {l c}
		\toprule
		Method                    & Log Bayes factor   \\
		\midrule
		BHM SDDR (classical)      & $0.951 \pm 0.012$  \\
		BHM SDDR (flows)          & $0.947 \pm 0.011$  \\ \midrule
		SBI learned harmonic mean & $1.093 \pm 0.014 $ \\
		SBI SDDR (classical)      & $0.905 \pm 0.018$  \\
		SBI SDDR (flows)          & $0.934 \pm 0.020$  \\
		\bottomrule
	\end{tabular}
	\caption{Field-level cosmology example comparing \wcdm\ vs. \lcdm. }
	\label{tab:sbi_lens}
\end{table}

\section{Conclusions} \label{sec:conclusions}

In this work we have developed a neural methodology to calculate the Savage-Dickey density ratio (SDDR) using normalizing flows. We have shown that our SDDR method scales to cases with many extra parameters involving high-dimensional marginal posterior spaces. We subsequently compared it to a classical approach of making use of normalized histograms, which is computationally challenging when there are many extra parameters. In low-dimensional marginal settings, both methods produce accurate and precise Bayes factor estimates.
As expected, when extending the analysis to higher-dimensional marginal settings such as the 4D marginal posterior toy Gaussian example presented in Sec. \ref{sec:gaussian_toy}, the classical method begins to struggle. The normalizing flow method however continues to provide accurate and precise calculations of the Bayes factor.
We further validated our normalizing flow methodology by producing consistent results with the classical approach and Bayes factors calculated with both nested sampling and the learned harmonic mean.

For a field-level inference setting, where the number of parameters is very large but the number of extra parameters of the super model is low, we leverage the SDDR to perform Bayesian model comparison for the first time.  Furthermore, we show that Bayes factors computed for BHM (Bayesian hierarchical model) and SBI (simulation-based inference) inference approaches are consistent, providing further validation that the SBI approach has extracted as much cosmological information from the field as the BHM approach.

Our neural method also requires very little computational resources to train and run, typically running within a few minutes on a single CPU, given the MCMC chains.
Making use of the neural SDDR with normalizing flows can approximately halve the computational cost of performing model comparison when working with nested models since one only needs to obtain posterior samples from the super model.
The code implementing the neural SDDR methodology with normalizing flows is included within the latest \texttt{harmonic} software package release.


\section*{Acknowledgements}
KL is supported by STFC (grant number ST/W001136/1) and the UK Space Agency (grant ST/X00208X/1). AP is supported by the UCL Centre for Doctoral Training in Data Intensive Science (STFC grant number ST/W00674X/1).   DP was supported by the SNF Sinergia grant CRSII5-193826 ``AstroSignals: A New Window on the Universe, with the New Generation of Large Radio-Astronomy Facilities''.  JDM is supported in part by STFC (grant number ST/W001136/1) and EPSRC (grant number EP/W007673/1).


\bibliographystyle{mnras}
\bibliography{bibliography}



\appendix

\section{SDDR Derivation} \label{sec:appendix}

What follows is a derivation of SDDR based on \citet{verde2013lack}. Suppose we have two models that are nested, where $M_1$ is the nested model and $M_2$ is the super model with \parsone\ common parameters and \parstwo\ extra parameters that extend the nested model to the super model. The Bayes factor is given by
\begin{eqnarray}
	\label{eq_ap:bf_sddr}
	\bayesfactor & = & \frac{\prob(\data | M_1)}{\prob(\data | M_2)} \\
	& = & \frac{
		\int \prob(\data | \parsone^{\prime}, M_1) \,
		\prob(\parsone^{\prime} | M_1) \,
		\diff \parsone^{\prime}
	}
	{
		\int \prob(\data | \parsone^{\prime\prime}, \parstwo^{\prime\prime}, M_2)  \,
		\prob(\parsone^{\prime\prime}, \parstwo^{\prime\prime} | M_2) \,
		\diff \parsone^{\prime\prime} \,
		\diff \parstwo^{\prime\prime}
	}\\
	&=& \frac{
		\int \prob(\data | \parsone^{\prime}, \parstwo_1,M_2)  \,
		\prob(\parsone^{\prime} | M_1) \,
		\diff \parsone^{\prime}
	}
	{
		\int \prob(\data | \parsone^{\prime\prime}, \parstwo^{\prime\prime},M_2) \,
		\prob(\parsone^{\prime\prime} | M_1) \,
		\prob( \parstwo^{\prime\prime} | M_2) \,
		\diff \parsone^{\prime\prime} \,
		\diff \parstwo^{\prime\prime}
	} \, ,
\end{eqnarray}
where in the second line we have used the fact that as the models are nested, we have $p(\data | \parsone, M_1) = \prob(\data | \parsone, \parstwo_1,M_2)$. In the third line we have used the fact that the prior distributions of \parsone\ are identical for $M_1$ and $M_2$ and that the prior distributions of \parstwo\ and \parsone\ are separable, i.e. $\prob(\parsone, \parstwo) = \prob(\parsone) \prob(\parstwo)$.

Multiplying both numerator and denominator by $\prob(\parstwo_1 | M_2)$ then gives
\begin{equation} \label{eq_ap:bf_sddr2}
	\bayesfactor  =  \frac{1}{\prob( \parstwo_1 | M_2)}
	\frac{
		\int \prob(\data | \parsone^{\prime}, \parstwo_1,M_2) \,
		\prob(\parsone^{\prime} | M_1) \,
		\prob( \parstwo_1 | M_2) \,
		\diff \parsone^{\prime}
	}
	{
		\int \prob(\data| \parsone^{\prime\prime}, \parstwo^{\prime\prime},M_2) \,
		\prob(\parsone^{\prime\prime} | M_1) \,
		\prob( \parstwo^{\prime\prime} | M_2) \,
		\diff \parsone^{\prime\prime} \,
		\diff \parstwo^{\prime\prime}
	} \, .
\end{equation}

\noindent To simplify this expression, we note that the normalized parameter posterior for $(\parsone, \parstwo)$ under the second model applicable to all values of $\parstwo$ is given by

\begin{equation}
	\prob(\parsone, \parstwo | \data, M_2)
	= \frac{\prob(\data | \parsone, \parstwo, M_2) \,
		\prob(\parsone | M_1) \,
		\prob(\parstwo | M_2)}
	{\int \prob(\data| \parsone^{\prime\prime}, \parstwo^{\prime\prime},M_2) \,
		\prob(\parsone^{\prime\prime} | M_1) \,
		\prob(\parstwo^{\prime\prime} | M_2) \,
		\diff \parsone^{\prime\prime} \,
		\diff \parstwo^{\prime\prime}}.
\end{equation}

\noindent Comparison with Eq. \eqref{eq_ap:bf_sddr2} and setting $\parstwo = \parstwo_1$ then yields

\begin{equation}
	\bayesfactor
	=\frac{
		\int \prob(\parsone^\prime, \parstwo_1 | \data, M_2) \,
		\diff \parsone^\prime
	}
	{ \prob( \parstwo_1 | M_2) }\, .
\end{equation}

We might notice that the integral in the numerator is performed only over a subset of parameters, and so is akin to marginalization, reducing to the marginalized posterior $\prob(\parstwo_1 | \data, M_2)$. Hence, the evidence ratio or Bayes factor between two nested models is given by

\begin{equation}
	\label{eq_ap:sddr}
	\bayesfactor
	= \frac{\prob(\parstwo_1 | \data, M_2)}{\prob(\parstwo_1 | M_2)} \, .
\end{equation}


\end{document}